**Title:**

Fast universal quantum control above the fault-tolerance threshold in silicon

**Authors:**

Akito Noiri[1,*], Kenta Takeda[1], Takashi Nakajima[1], Takashi Kobayashi[2], Amir Sammak[3], Giordano Scappucci[4], and Seigo Tarucha[1,2,*]

**Affiliations:**

[1]*RIKEN, Center for Emergent Matter Science (CEMS), Wako-shi, Saitama 351-0198, Japan*
[2]*RIKEN, Center for Quantum Computing (RQC), Wako-shi, Saitama 351-0198, Japan*
[3]*QuTech and Netherlands Organisation for Applied Scientific Research (TNO), Stieltjesweg 1, 2628 CK Delft, Netherlands*
[4]*QuTech and Kavli Institute of Nanoscience, Delft University of Technology, Lorentzweg 1, 2628 CJ Delft, Netherlands*

*e-mail: akito.noiri@riken.jp or tarucha@riken.jp

**Summary paragraph:**

Fault-tolerant quantum computers which can solve hard problems rely on quantum error correction[1]. One of the most promising error correction codes is the surface code[2], which requires universal gate fidelities exceeding the error correction threshold of 99 per cent[3]. Among many qubit platforms, only superconducting circuits[4], trapped ions[5], and nitrogen-vacancy centers in diamond[6] have delivered those requirements. Electron spin qubits in silicon[7–15] are particularly promising for a large-scale quantum computer due to their nanofabrication capability, but the two-qubit gate fidelity has been limited to 98 per cent due to the slow operation[16]. Here we demonstrate a two-qubit gate fidelity of 99.5 per cent, along with single-qubit gate fidelities of 99.8 per cent, in silicon spin qubits by fast electrical control using a micromagnet-induced gradient field and a tunable two-qubit coupling. We identify the condition of qubit rotation speed and coupling strength where we robustly achieve high-fidelity gates. We realize Deutsch-Jozsa and Grover search algorithms with high success rates using our universal gate set. Our results demonstrate the universal gate fidelity beyond the fault-tolerance threshold and pave the way for scalable silicon quantum computers.

**Main text:**

Electron spins in silicon quantum dots are an attractive platform of a quantum computer with a long coherence time[7–9], capability of high-temperature operation[10,11], and potential scalability[12–15]. Single-qubit gate fidelity higher than the fault-tolerance threshold is now routinely achieved[7,8,17]. Two-qubit gate fidelity, on the other hand, still remains 98%[16], below the threshold because of complexity of operation and/or slow operation compared to the coherence time[16,18–20]. Native two-qubit gates for spin qubits include $\sqrt{\text{SWAP}}$[21–24], controlled-phase[9,18,20], and controlled-rotation (CROT)[18,19], all relying on the exchange coupling. Rapid control of exchange coupling by gate voltage pulses enables $\sqrt{\text{SWAP}}$[21–24] and controlled-phase gates[9,18,20] at the cost of requiring high-bandwidth and precise pulse engineering which obstructs a high-fidelity gate. In contrast, a CROT[18,19] can be implemented with less demanding pulse engineering in a fixed coupling[16]. With additional adjustments of single-qubit phases, a controlled-NOT (CNOT) gate with fidelity 98% is demonstrated. Since the fidelity is mostly limited by dephasing[16], it is crucial to mitigate the dephasing effect by a faster operation to go beyond the fault-tolerance threshold. Furthermore, reliable and efficient tuning strategy of the high-fidelity gate is desired for scaling up the silicon spin qubits.

In this work, we realize a CNOT gate fidelity of 99.5% with a gate time of 103 ns in an isotopically enriched silicon quantum dot array. Our device satisfies three key elements to achieve this. First, the exchange coupling $hJ$ is widely controllable to make it large enough at the charge-symmetry point where the effect of charge noise during a fast operation is suppressed[19,22]. Here $h$ is the plank constant. Second, the Zeeman energy difference between the qubits $hdE_\text{Z}$ induced by a micromagnet is also large enough to allow a large $hJ$. Finally, the CROT gates are implemented by fast electric-dipole spin resonance (EDSR) controls of single spins driven in the slanting magnetic field induced by the micromagnet. These device features enable us to assess the single- and two-qubit gate performances over a wide range of parameters that were not accessible in the previous work[16]. The comprehensive study of the gate performances reveal that they mainly depend on the gate speed, from which we identify the gate condition where the CNOT gate fidelity higher than 99% is robustly achieved. In the same gate condition, single-qubit gate fidelities reach 99.8% for both qubits. Utilizing the high-fidelity universal quantum control, we implement the two-qubit Deutsch–Jozsa algorithm[25] and the Grover search algorithm[26] with the success rates of 96-97%. These results demonstrate the universal quantum control fidelity exceeding the surface code error correction threshold, showing that high-fidelity quantum processing is feasible in silicon spin qubits.

The device is a linearly coupled triple quantum dot fabricated on an isotopically enriched silicon/silicon-germanium heterostructure (Methods). Three layers of aluminum gates create confinement potentials to define the quantum dots[15] (Fig. 1a). The center (right) quantum dot has an electron which is operated as a qubit $Q_1$ ($Q_2$) while the left dot is not formed but used as an extension of the left reservoir. On top of the aluminum gates, a cobalt micromagnet is fabricated to generate a magnetic field gradient required for the fast EDSR control of both qubits[27] and also induce $dE_Z$ for the CROT gates.

Figure 1b shows a typical charge stability diagram around the charge configurations to define the qubits. Each qubit is sequentially initialized and measured in a single-shot manner by energy-selective electron tunneling between the quantum dots and neighboring reservoirs[28,29] at the gate voltage conditions shown by the white circles in Fig. 1b. The qubits are manipulated around the charge-symmetry point in the (1,1) charge state (Extended Data Fig. 1) shown by the white square to suppress charge-noise sensitivity during operations[19,22]. Here, (c, r) denotes the number of electrons in the center (c) and right (r) dots.

This two-qubit system is capable of implementing the universal gate set. Under an EDSR control with a Rabi frequency of $f_R$, a microwave frequency of $f_{MW}$ and its phase $\phi$, a Hamiltonian for the two-qubit system in the basis of $|\uparrow\uparrow\rangle$, $|\widetilde{\uparrow\downarrow}\rangle$, $|\widetilde{\downarrow\uparrow}\rangle$, and $|\downarrow\downarrow\rangle$ can be approximated as[16]

$$H = \frac{h}{2}\begin{pmatrix} 2E_Z & \Omega & \Omega & 0 \\ \Omega^* & -d\tilde{E}_Z - J & 0 & \Omega \\ \Omega^* & 0 & d\tilde{E}_Z - J & \Omega \\ 0 & \Omega^* & \Omega^* & -2E_Z \end{pmatrix},$$

where $hE_Z$ is the average Zeeman energy, $hd\tilde{E}_Z = h\sqrt{dE_Z^2 + J^2}$ is the effective Zeeman energy difference between the qubits, and $\Omega = f_R e^{i2\pi f_{MW}t + i\phi}$ is the EDSR driving. The tilde indicates the hybridization of the spin eigenstates $|\uparrow\downarrow\rangle$ and $|\downarrow\uparrow\rangle$ due to the exchange coupling. Then each EDSR frequency is given by $f_{m,\sigma} = E_Z \pm (d\tilde{E}_Z \pm J)/2$ (Fig. 1c) where m is the index of the target qubit $Q_m$ (m = 1 or 2) and $\sigma$ is the control qubit state $|\uparrow\rangle$ or $|\downarrow\rangle$. When the separations of $f_{m,\sigma}$ are larger than the width of each EDSR spectrum, a CROT gate can be implemented by driving one of the EDSR transitions[16,19] (Fig. 1d). An in-plane external magnetic field of $B_{ext} = 0.408$ T results in $E_Z \sim 15700$ MHz. The micromagnet induces $dE_Z \sim 300$ MHz and we can control $J$ from a few MHz to tens of MHz by varying the barrier gate voltage. While a larger $f_R$ is desired for high-fidelity CROT gate, it also results in unwanted rotation of the off-resonant states. To cancel this unwanted rotation in both π and π/2 CROT gates (Methods)[16,30], hereafter we use $f_R = J/\sqrt{15}$ unless specifically noted.

Two important characteristics that can influence both single- and two-qubit gate performances are the dephasing and the decay of Rabi oscillation during the gate time. Figure 1e shows $J$ (and $f_R$) dependence of the dephasing times $T_{2,m,\sigma}^*$ measured for each transition (see also Extended Data Fig. 2d-f for detail). We find that $T_{2,m,\sigma}^*$ is almost constant in the measured range of $J$ since they are mostly limited by single-qubit frequency noise rather than the fluctuation in $J$ as corroborated by noise measurement (Extended Data Fig. 5a). This implies that increasing $f_R$ with keeping $J = f_R\sqrt{15}$ is favorable to suppress the dephasing effect as larger $J$ does not introduce extra dephasing. In contrast, we find that the Rabi decay depends on $f_R$. Figure 1f shows the $f_R$ dependence of Rabi decay $D_{m,\sigma}$ during a π/2 CROT (see Extended Data Fig. 2j-l for detail). One can expect that the coherence-limited single- and two-qubit gate performances[16,17,31] are improved with decreasing $D_{m,\sigma}$[8,32]. At small ($f_R \leq 2$ MHz) Rabi frequencies, $D_{m,\sigma}$ decreases with increasing $f_R$ since the effect of dephasing is suppressed. On the other hand, $D_{m,\sigma}$ increases with $f_R$ above approximately 5 MHz since the Rabi decay becomes faster possibly due to heating and/or population leakage[8,32]. In between the two regimes, $D_{m,\sigma}$ is minimized. This simple observation indicates that the best performance of single- and two-qubit gates should be obtained at $f_R \sim 5$ MHz.

Then, we measure basic qubit properties and assess single- and two-qubit gate fidelities at $f_R = 4.867$ MHz and $J = 18.85$ MHz. The spin relaxation times for both qubits are much longer than the maximum operation time of 100 μs (Extended Data Fig. 2c) and therefore the spin relaxation effect is negligible in the gate performances. The dephasing times $T^*_{2,m,\sigma}$ are several μs (Fig. 1e) and they are enhanced by the Hahn echo sequence up to $\sim 30$ μs (Extended Data Fig. 2i). Single-qubit gate fidelities are characterized by the Clifford-based randomized benchmarking (Fig. 2b and Methods)[33]. In this system, a single-qubit gate is constructed from two CROT gates (Fig. 2a). We obtain primitive gate fidelities of $F_{p,1} = 99.84 \pm 0.004\%$ for $Q_1$ and $F_{p,2} = 99.84 \pm 0.004\%$ for $Q_2$ (Fig. 2c). The two-qubit gate fidelity is also characterized by the Clifford-based two-qubit randomized benchmarking. All of the Clifford gates are constructed from the primitive gates shown in Fig. 2d, another set of gates where the roles of $Q_1$ and $Q_2$ are swapped and single-qubit phase gates acting on each qubit[16]. Using the quantum circuit shown in Fig. 2e, we obtain a Clifford gate fidelity $F_C = 98.67 \pm 0.01\%$ which corresponds to a primitive gate fidelity $F_p = 99.48 \pm 0.004\%$ as shown in Fig. 2g (Methods). Since all of the primitive gates similarly comprise two π/2 CROT gates[16], each primitive gate including the CNOT gate should have similar gate fidelity. To confirm this, we directly assess the CNOT gate fidelity $F_{CNOT}$ by the interleaved randomized benchmarking[4,34]. By comparing the sequence fidelity decay using the quantum circuit shown in Fig. 2e, f, we obtain $F_{CNOT} = 99.51 \pm 0.02\%$ (Fig. 2g and Methods) which agrees with $F_p$.

Next, we measure the impact of $f_R$ on the single- and two-qubit gate performances to study robustness of the high-fidelity gates. Figure 3a shows the $f_R$ dependence of single-qubit primitive gate fidelities $F_{p,m}$. The best performance of $F_{p,m} \sim 99.8\%$ is obtained at $f_R = 2$-$5$ MHz, in agreement with $f_R$ dependence of the Rabi decay (Fig. 1f). The two-qubit primitive gate fidelity $F_p$ also depends on $f_R$ as shown in Fig. 3b. For small $f_R$ ($\leq 2.8$ MHz), $F_p$ is below 99% and mostly limited by dephasing[16]. In this regime, $F_p$ is much lower than $F_{p,m}$ since the dephasing effect mainly affects the control qubit that is left idle while the target qubit is driven in CROT gates. Therefore, suppressing the dephasing effect is more important in the two-qubit gates. By increasing $f_R$, the dephasing effect can be suppressed and we obtain $F_p$ above 99%. By further increasing $f_R$, $F_p$ sharply drops due to the fast Rabi decay. As expected from $f_R$ dependence of the dephasing time and the Rabi decay (Fig. 1e, f), we obtain the best values of $F_p \sim 99.5\%$ at $f_R = 4\text{-}5$ MHz. To consider the limiting factor of $F_p$ in this condition, we simulate the effect of dephasing (Methods and Extended Data Fig. 5a, b). We find that the infidelity due to dephasing is only 0.1% and therefore it is no longer the main limiting factor. The effect of Rabi decay is also small as indicated by the high-fidelitiy (99.84%) single-qubit gates (Fig. 2c). The remaining infidelity could originate from pulse calibration errors and long-term fluctuations in the device condition. These results indicate that the optimal gate condition is efficiently searched only from simple measurements of dephasing time and Rabi decay, which will be useful to tune up a large qubit array.

Finally, we implement the two-qubit Deutsch–Jozsa algorithm[25] and Grover search algorithm[26] to demonstrate the feasibility of high-fidelity quantum processing. The Deutsch–Jozsa algorithm (Fig. 4a) determines whether an unknown function $f_i(x)$ mapping a single-bit input $x \in \{0,1\} = \{|\uparrow\rangle, |\downarrow\rangle\}$ to a single-bit output is constant ($f_0(x) = 1$, $f_1(x) = 0$) or balanced ($f_2(x) = x$, $f_3(x) = 1 - x$) by a single call of the function. The Grover search algorithm (Fig. 4b) finds the unique input two-bit string $x_0 = ij$ (i, j $\in \{0,1\}$) of a function $f_{ij}(x)$ which outputs 0 for $x \neq x_0$ but 1 for $x = x_0$ by a single call of the function. Figure 4c-f (4g-j) shows the real part of the density matrix (Methods) measured at each stage for $f_2$ ($f_{11}$). All through the processing, the state fidelity compared to the ideal state is kept high (> 96%), which surpasses those previously obtained in silicon spin qubits[18,35]. We also obtain similar output state fidelities for the other functions (Extended Data Fig. 8). These results demonstrate that high-fidelity quantum processing is feasible in silicon spin qubits.

In conclusion, we demonstrate single- and two-qubit primitive gate fidelities of 99.8% and 99.5%, respectively, which are beyond the surface code error correction threshold[3]. Micromagnet-induced gradient field and tunable exchange coupling allow us to assess the single- and two-qubit gate fidelities with a variety of gate conditions and reveal a relationship between them. We identify that the Rabi frequency for single-qubit rotation influences both single- and two-qubit gate fidelities. We find a range of Rabi frequency where we robustly achieve the two-qubit primitive and CNOT gate fidelities higher than 99%. The demonstrated universal quantum gate set allows us to implement two-qubit quantum algorithms with high fidelities. Our results are an important step toward realizing fault-tolerant quantum computation in silicon spin qubits.

During the completion of this manuscript, we became aware of related experiments that demonstrate universal quantum control fidelity exceeding the fault-tolerance threshold in two electron spin qubits[36] and two nuclear spin qubits[37] in silicon.

**Figures:**

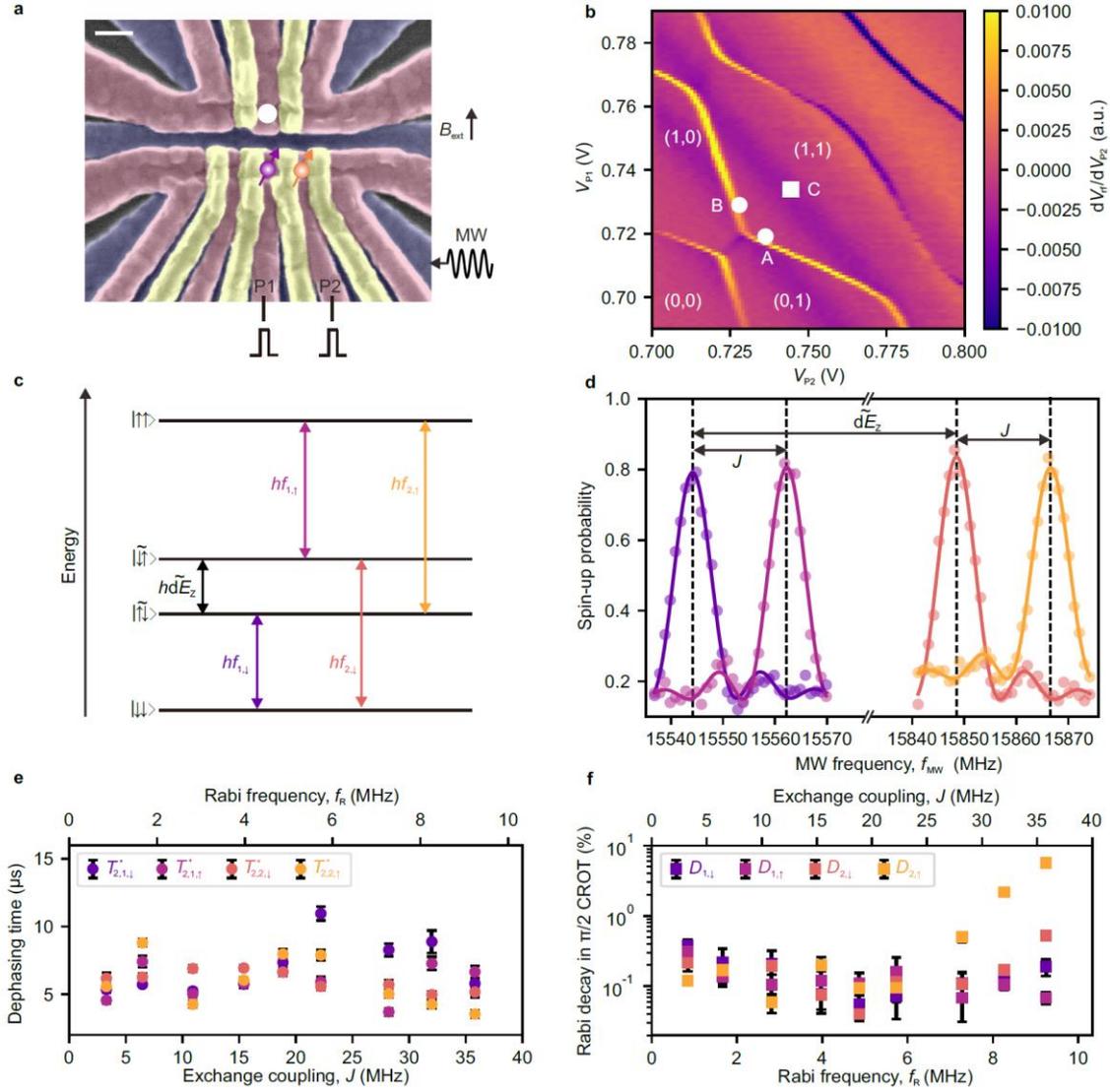

**Figure 1. Two-qubit system. a,** False color scanning microscope image of a device identical to the one measured. The qubits are located underneath the P1 and P2 gate electrodes. The white circle shows a charge sensor quantum dot embedded in a radio-frequency tank circuit[38,39]. The white scale bar indicates 100 nm. **b,** Charge stability diagram around the operation condition. Initialization and measurement for $Q_1$ ($Q_2$) is performed at the white circle labeled A (B). Qubits manipulation is performed at the charge-symmetry point shown in the white square labeled C. **c,** Energy diagram of the two-qubit system. Each colored arrow shows the state transition driven by EDSR with the microwave frequency $f_{1,\downarrow}$, $f_{1,\uparrow}$, $f_{2,\downarrow}$, and $f_{2,\uparrow}$. **d,** EDSR spectra for $Q_1$ when $Q_2$ is spin-down (purple) and -up (magenta) and for $Q_2$ when $Q_1$ is spin-down (orange) and -up (yellow). **e,** $J$ (and $f_R$) dependence of the dephasing times (Extended Data Fig. 2d-f). We choose the charge-symmetry point

as the operation condition and control $J$ by modifying the tunnel coupling between the quantum dots. The errors represent the estimated standard errors for the best-fit values. **f**, $f_R$ (and $J$) dependence of the Rabi decay during a $\pi/2$ CROT obtained from the Rabi decay curves (Extended Data Fig. 2j-l). The errors represent the estimated standard errors for the best-fit values.

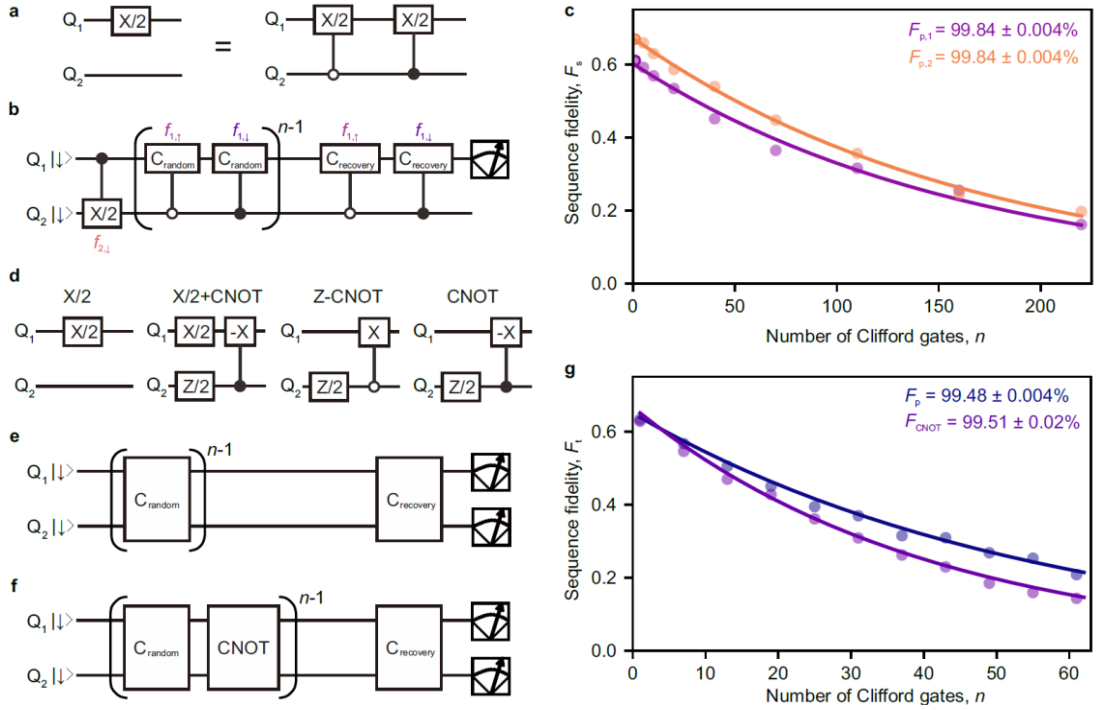

**Figure 2. Characterization of universal quantum control performances by randomized benchmarking. a,** Example of a single-qubit gate constructed from two CROT gates to make it unconditional spin rotation. **b,** Quantum circuit to measure the single-qubit gate fidelity of $Q_1$ by the Clifford-based randomized benchmarking. As the spin rotation under finite exchange coupling is a CROT, we synthesize our unconditional single-qubit rotations to assess the averaged performance of single-qubit gate of $Q_1$ with various $Q_2$ states. The roles of $Q_1$ and $Q_2$ are swapped to measure the single-qubit gate fidelity of $Q_2$. **c,** Single-qubit Clifford-based randomized benchmarking for $Q_1$ (purple) and $Q_2$ (orange). The uncertainty in the gate fidelities are obtained by a Monte Carlo method[4]. **d,** Quantum circuits for two-qubit primitive gates that rotate $Q_1$. CNOT and zero-CNOT (Z-CNOT) gates flip the target qubit $Q_1$ when the control qubit $Q_2$ is spin-down and -up[16]. We construct our Clifford gates so that the number of primitive gates per one Clifford gate is minimum, which results in 2.57 primitive gates per one Clifford gate on average[16]. Since the single-qubit phase gates are implemented by changing the reference frame of all subsequent microwave pulses in the software, we do not include them in the primitive gate count. **e,** Quantum circuit for two-qubit randomized benchmarking to measure the Clifford gate fidelity and the primitive gate fidelity $F_p$. **f,** Quantum circuit for interleaved randomized benchmarking to measure the CNOT gate fidelity $F_{CNOT}$. **g,** Results of the two-qubit Clifford-based randomized benchmarking. The uncertainty in gate fidelities are obtained by a Monte Carlo method[4].

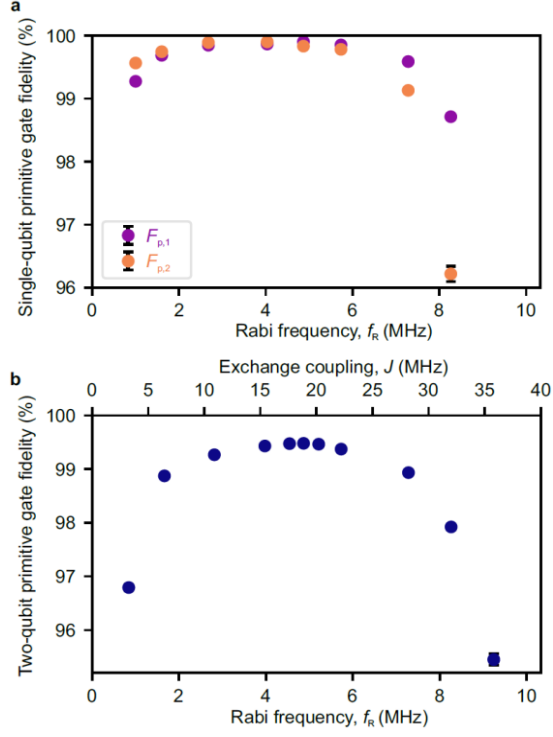

**Figure 3. Rabi frequency dependence of single- and two-qubit primitive gate fidelities. a,** $f_R$ dependence of the single-qubit primitive gate fidelity extracted from single-tone single-qubit primitive gate fidelities (Extended Data Fig. 3d) as $F_{p,m} = F_{p,m,\downarrow} F_{p,m,\uparrow}$. Here we measure single-tone single-qubit primitive gate fidelities to assess the impact of $f_R$ on the single-qubit gate performance without involving the effect of $J$ (Extended Data Fig. 3d). **b,** $f_R$ and $J$ dependence of the two-qubit primitive gate fidelity. $f_R = J/\sqrt{15}$ is always satisfied. When the Rabi frequency is in between 2.8 and 5.7 MHz, both dephasing and Rabi decay effects are suppressed and the two-qubit primitive gate fidelity exceeds the fault-tolerance threshold. The uncertainty in the fidelities are obtained by a Monte Carlo method[4].

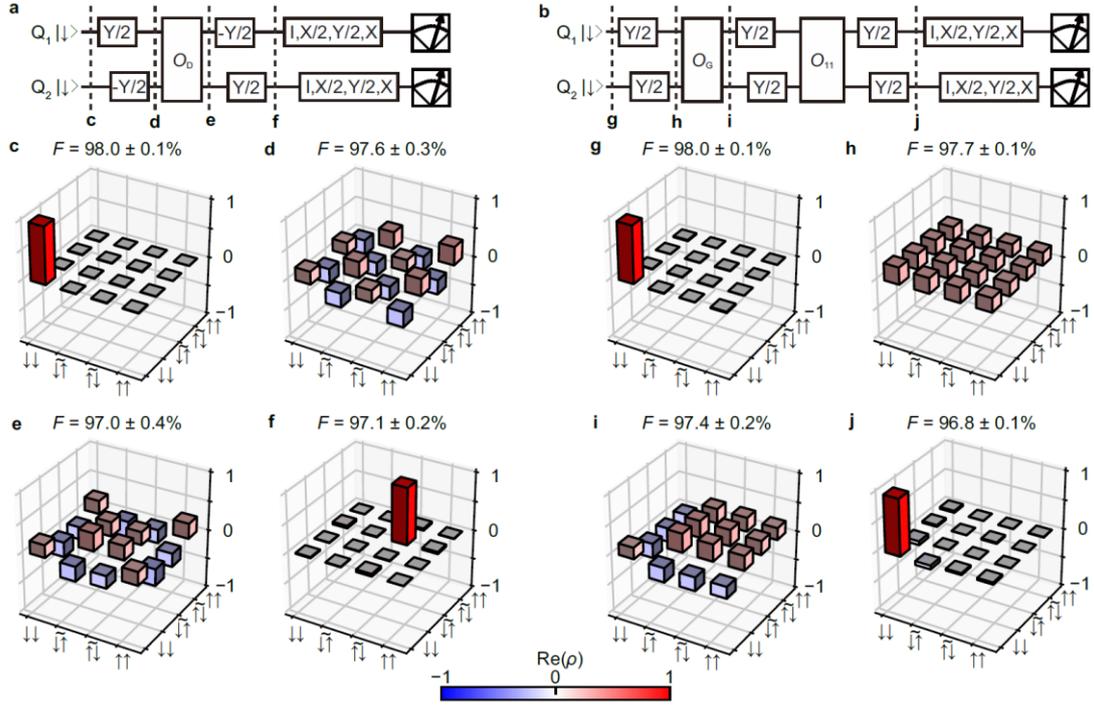

**Figure 4. Two-qubit quantum processing. a,** Quantum circuit for the two-qubit Deutsch–Jozsa algorithm[18]. $f_i(x)$ is implemented by an oracle $O_D = I_2$ for $f_0$, $X_2$ for $f_1$, $Z - CNOT_2$ for $f_2$, and $CNOT_2$ for $f_3$ where the subscript 2 after gates indicates that the target qubit is $Q_2$. **b,** Quantum circuit for the two-qubit Grover search algorithm[18]. $f_{ij}(x)$ is implemented by an oracle $O_G = O_{11} = (Y_2/2)(CNOT_2)(-Y_2/2)$ for $f_{11}$, $O_{10} = (Y_1/2)(Z - CNOT_1)(-Y_1/2)$ for $f_{10}$, $O_{01} = (-Y_1/2)(CNOT_1)(Y_1/2)$ for $f_{01}$, and $O_{00} = (-Y_2/2)(Z - CNOT_2)(Y_2/2)$ for $f_{00}$. **c-f,** Real part of the measured density matrix (Methods) for $f_2$ after initialization (**c**), preparation of the input state (**d**), application of the oracle (**e**), and the completion of the processing (**f**). **i-k,** Real part of the measured density matrix for $f_{11}$ at each stage shown in **b**. The absolute values of the matrix elements for the imaginary parts are less than 0.028 (**c**), 0.046 (**d**), 0.051 (**e**), 0.050 (**f**), 0.013 (**g**), 0.072 (**h**), 0.081 (**i**), and 0.079 (**j**). The uncertainty in the state fidelities $F$ are obtained by a Monte Carlo method[16,18,40].

**Methods:**

**Measurement setup.**

The sample is cooled down in a dry dilution refrigerator (Oxford Instruments Triton) with a base temperature of $\sim 20$ mK. The electron temperature is $\sim 60$ mK. The dc gate voltages are supplied by a 24-channel digital-to-analog converter (QDevil ApS QDAC), which is low-pass filtered at a cutoff frequency of 800 Hz. The voltage pulses applied to the P1 and P2 gate electrodes are generated by an arbitrary waveform generator (Tektronix AWG5014C). The EDSR microwave pulses are generated using an I/Q modulated signal generator (Anapico APMS20G with a Marki microwave MLIQ-0218 I/Q mixer) and applied to the bottom screening gate. The I/Q modulation signals are generated by another arbitrary waveform generator (Tektronix AWG70002A) triggered by the arbitrary waveform generator used for generating the gate voltage pulses. The microwave signals are sideband-modulated by frequencies ranging from -240 to 180 MHz from the baseband frequency in order to avoid the unintentional spin rotation due to leakage (the typical isolation is $\sim 50$ dBc after calibration of the I/Q imbalances and the dc offsets) as well as switch the microwave frequencies rapidly. During the initialization and measurement stages, additional pulse modulations are used to provide further isolation of the microwave signals. The Rabi frequencies of single-qubit rotations are controlled by the amplitudes of I/Q modulation signals.

**Sample fabrication.**

The quantum dots are defined at the isotopically enriched silicon quantum well (residual $^{29}$silicon concentration of 800 parts per million) 50 nm below the wafer surface. Three layers of overlapping aluminum gates are fabricated by electron-beam lithography and lift-off processes[15]. Each layer is insulated by the thin native aluminum oxide. The micromagnet made of a stack of titanium and cobalt films with thicknesses of 5 and 250 nm is placed on top of the overlapping gates with a 30 nm thick insulating layer (aluminum oxide grown by an atomic layer deposition) in between. The micromagnet design is similar to those in previous reports[8,19,27,40].

**Sequence fidelity and gate fidelity extraction in randomized benchmarking.**

The sequence fidelity of single-qubit randomized benchmarking is obtained by the following procedure. According to the standard randomized benchmarking protocol, we measure spin-up probability as a function of the number of Clifford gates $n$. Then the spin-up probability $P_\uparrow$ follows $P_\uparrow(n) = A_s p_s^n + C_s$, where $p_s$ is the depolarizing parameter, $A_s$ and $C_s$ are the constants to absolve the state preparation and measurement errors. Here, the recovery Clifford gate is chosen so that the final ideal state is spin-up. We also obtain another data set where the final ideal state is spin-down by choosing different recovery gates. In this case, the spin-up probability $P'_\uparrow$ follows $P'_\uparrow(n) = B_s p_s^n + C_s$, where $B_s$ is the constant determined by state preparation and measurement errors. Then the sequence fidelity $F_s(n)$ is obtained from $F_s(n) = P_\uparrow(n) - P'_\uparrow(n) = (A_s - B_s)p_s^n$, eliminating the uncertainty of determining $C_s$[7,18,40,41]. Here the fitting parameter $A_s - B_s$ absorbs the state preparation and measurement errors. We average 16 random sequences, each of which are repeated 400 times to measure $F_s(n)$. The Clifford gate fidelity $F_{C,m}$ is obtained by $F_{C,m} = (1 + p_s)/2$ where m is the qubit number 1 or 2. Since a Clifford gate contains 1.875 primitive gates on average, we extract the primitive gate fidelity $F_{p,m}$ as $F_{p,m} = 1 - (1 - F_{C,m})/1.875$.

Similarly, the sequence fidelity of two-qubit randomized benchmarking is extracted by the following procedure. We measure $P_{\uparrow\uparrow}(n) = A_t p_t^n + C_t$ as a function of $n$ where $P_{\uparrow\uparrow}$ is the joint probability of spin-up in both qubits, $p_t$ is the depolarizing parameter, $A_t$ and $C_t$ are the constants to absolve the state preparation and measurement errors. Here, the recovery Clifford gate is chosen so that the final ideal state is spin-up for both qubits. We also measure another data set where the final ideal state is spin-down for both qubits and obtain $P'_{\uparrow\uparrow}(n) = B_t p_t^n + C_t$. $B_t$ is the constant determined by state preparation and measurement errors. Then the sequence fidelity $F_t(n)$ is extracted from $F_t(n) = P_{\uparrow\uparrow}(n) - P'_{\uparrow\uparrow}(n) = (A_t - B_t)p_t^n$. We average 60 random sequences each of which are repeated 400 times to measure $F_t(n)$. The two-qubit Clifford gate fidelity is obtained by $F_C = (1 + 3p_t)/4$. Since a Clifford gate contains 2.57 primitive gates on average, we extract the primitive gate fidelity $F_p$ as $F_p = 1 - (1 - F_C)/2.57$. The obtained gate fidelity using this protocol agrees with that obtained using the standard protocol[4,16] (only measures $P_{\uparrow\uparrow} = A_t p_t^n + C_t$) as shown in Extended Data Fig. 4.

Fidelity of the CNOT gate is obtained as follows[4]. We first measure $F_t(n)$ by applying random Clifford gates (Fig. 2e) and obtain the depolarizing parameter $p_{\text{ref}}$ as a reference. We also measure $F_t(n)$ by applying the CNOT gate between each random Clifford gates (Fig. 2f) and obtain the depolarizing parameter $p_{\text{CNOT}}$. Then we extract the CNOT gate fidelity as $F_{\text{CNOT}} = (1 + 3p_{\text{CNOT}}/p_{\text{ref}})/4$.

The errors of the gate fidelities are obtained by a Monte Carlo method[4]. We fit the resulting fidelity distribution by the Gaussian distribution and extract its standard deviation.

**Estimation of resonance frequencies fluctuations.**

Time dependence of the resonance frequencies $f_{1,\downarrow}$, $f_{1,\uparrow}$, $f_{2,\downarrow}$, and $f_{2,\uparrow}$ are extracted from repeated Ramsey fringe measurements. We sequentially measure Ramsey fringes for Q$_1$ (Q$_2$) when Q$_2$ (Q$_1$) is spin-down and -up by changing the evolution time from 0.04 μs to 4.0 μs with 0.04 μs step. Then we estimate each resonance frequency from single record of Ramsey fringe by Bayesian estimation[31,42]. This single cycle takes 1.706 s. We repeat the measurement 10,000 cycles and extract time dependence of $f_{1,\downarrow}$, $f_{1,\uparrow}$, $f_{2,\downarrow}$, and $f_{2,\uparrow}$. The fluctuation of $J/2$ ($\Delta J/2 = (\Delta f_{1,\uparrow} - \Delta f_{1,\downarrow})/2$), single-qubit frequencies of Q$_1$ ($\Delta f_1 = (\Delta f_{1,\uparrow} + \Delta f_{1,\downarrow})/2$), and Q$_2$ ($\Delta f_2 = (\Delta f_{2,\uparrow} + \Delta f_{2,\downarrow})/2$) are shown in Extended Data Fig. 5a. Since the fluctuation of each resonance frequency is $\Delta f_{m,\sigma} = \Delta f_m \pm \Delta J/2$ and $\Delta f_m$ is larger than $\Delta J/2$, the dephasing times are mostly limited by the noise in single-qubit frequencies rather than that of $J$ and therefore $J$ does not have a significant impact on the dephasing times as shown in Fig. 1e.

**Theoretical description of controlled-rotation.**

To understand the time evolution of the system under EDSR control, it is simpler to consider in a time dependent rotating frame $R = \text{diag}\,(e^{-i2\pi E_Z t}, e^{-i\pi(-d\tilde{E}_Z - J)t}, e^{-i\pi(d\tilde{E}_Z - J)t}, e^{i2\pi E_Z t})$. Then the Hamiltonian is described by $H_R(t) = RHR^\dagger - \frac{ih}{2\pi}\frac{\partial R}{\partial t}R^\dagger$

$$= \frac{h}{2}\begin{pmatrix} 0 & f_R e^{-i2\pi(f_{2,\uparrow} - f_{MW})t + i\phi} & f_R e^{-i2\pi(f_{1,\uparrow} - f_{MW})t + i\phi} & 0 \\ f_R e^{i2\pi(f_{2,\uparrow} - f_{MW})t - i\phi} & 0 & 0 & f_R e^{-i2\pi(f_{1,\downarrow} - f_{MW})t + i\phi} \\ f_R e^{i2\pi(f_{1,\uparrow} - f_{MW})t - i\phi} & 0 & 0 & f_R e^{-i2\pi(f_{2,\downarrow} - f_{MW})t + i\phi} \\ 0 & f_R e^{i2\pi(f_{1,\downarrow} - f_{MW})t - i\phi} & f_R e^{i2\pi(f_{2,\downarrow} - f_{MW})t - i\phi} & 0 \end{pmatrix}.$$

The off-diagonal terms result in a CROT gate by choosing one of the resonance frequencies $f_{1,\downarrow}$, $f_{1,\uparrow}$, $f_{2,\downarrow}$, and $f_{2,\uparrow}$. Here the effect of oscillating terms with an oscillation frequency much larger than $f_R$ is averaged out during the $\pi/2$ CROT time $t_{hp} = 1/(4f_R)$. When $J$ is comparable to $f_R$, the terms oscillating with a frequency of $J$ results in unwanted off-resonant rotation of the target qubit. The off-resonant rotation follows the Hamiltonian $\frac{h}{2}\begin{pmatrix} \pm J & f_R \\ f_R & \mp J \end{pmatrix}$ and therefore the target qubit rotates along a tilted axis with an effective Rabi frequency $\tilde{f}_R = \sqrt{f_R^2 + J^2}$. To suppress the effect, $\tilde{f}_R t_{hp}$ must be an integer, and therefore we use the Rabi frequency such that $f_R = J/\sqrt{16k^2 - 1}$ where k is an integer[16,30].

**Simulation of two-qubit gate infidelity by quasi-static noise in resonance frequencies.**

We simulate the effect of the resonance frequency noise[16] on the two-qubit primitive gate fidelity. We assume the noise is quasi-static in the single measurement of 100 μs but changes in between the measurements. We use the measured time dependence of $\Delta J$, $\Delta d\tilde{E}_Z = \Delta f_2 - \Delta f_1$, and $\Delta E_z = (\Delta f_1 + \Delta f_2)/2$ (Extended Data Fig. 5a) to simulate the effect. We calculate π/2 CROT operators by $U_{CROT} = \prod_{k=0}^{k=N} e^{-i2\pi(H_R(k\Delta t) + \Delta H_R)\Delta t/h}$ where $\Delta H_R = \text{diag}(2h\Delta E_Z, -h\Delta d\tilde{E}_Z - h\Delta J, h\Delta d\tilde{E}_Z - h\Delta J, -2h\Delta E_Z)/2$, $\Delta t = t_{hp}/N$, and N is a large integer (N = 1000 in our calculation) and obtain operators of the primitive gates. Then we calculate the probability of the ideal final state as a function of the number of randomly chosen Clifford gates. We average 60 random sequences, each of which is repeated 100 times with different $\Delta J$, $\Delta d\tilde{E}_Z$, and $\Delta E_z$. Then we extract the two-qubit primitive gate infidelity. In the simulation, $J$ is fixed at $\sqrt{15}f_R$. $f_R$ dependence of the infidelity is shown in

Extended Data Fig. 5b. Around the optimal gate condition $f_R = 4\text{-}5$ MHz, the infidelity is only $\sim 0.1\%$.

Future works will include a CNOT gate with pulsed exchange control[19] to make the system suitable for scaling up. In addition to switching the exchange coupling, this requires an additional idle time of $(2 - \sqrt{15}/2)/J$ in all of the primitive gates to make the total gate time $2/J$ to remove unwanted controlled-phase accumulation during the exchange pulse[19,30]. We simulate this case as shown in Extended Data Fig. 5c. Around $f_R = 4\text{-}5$ MHz, the infidelity caused by the additional idle time is less than $0.1\%$. Therefore, we anticipate a CNOT gate fidelity higher than $99\%$ with exchange pulses is within reach.

**State tomography.**

First, we remove the measurement error from the measured joint probabilities $P_M = (P_{\downarrow\downarrow}, P_{\downarrow\uparrow}, P_{\uparrow\downarrow}, P_{\uparrow\uparrow})$ and obtain the joint probabilities $P = (P_{\downarrow\downarrow}, P_{\widetilde{\downarrow\uparrow}}, P_{\widetilde{\uparrow\downarrow}}, P_{\uparrow\uparrow})$ which is used to extract its density matrix. To do this, we measure a readout correction matrix $C$ as shown in Extended Data Fig. 7. Here, four computational basis states ($|\uparrow\uparrow\rangle$, $|\widetilde{\uparrow\downarrow}\rangle$, $|\widetilde{\downarrow\uparrow}\rangle$, and $|\downarrow\downarrow\rangle$) are prepared and joint probabilities are measured[43]. Then, $P$ is obtained such that $P = C^{-1}P_M$.

Next, we perform maximum likelihood estimation to make the density matrix physical[16,18,19,40]. A physical density matrix $\rho$ can be described using a complex lower triangular matrix having real diagonal elements $T$ as $\rho = TT^\dagger/\text{Tr}(T^\dagger T)$. Then we minimize the cost function

$$C(\mathbf{t}) = \sum_{\nu=1}^{16} \frac{(\langle\psi_\nu|\rho(\mathbf{t})|\psi_\nu\rangle - P_\nu)^2}{2\langle\psi_\nu|\rho(\mathbf{t})|\psi_\nu\rangle},$$

where $\mathbf{t} = (t_1, t_2, \ldots, t_{16})$ is the real parameters of $T$, $P_\nu$ is the probability projected at a state $|\psi_\nu\rangle$ obtained by averaging measurement results of 10,000 shots with measurement error correction. To extract $\mathbf{t}$, 16 combinations of $(I, X/2, Y/2, X)$ pre-rotations acting on Q1 and Q2 are used[4,18]. The uncertainty of the state fidelity is obtained by a Monte Carlo method assuming that the measured single-shot probabilities follow multinomial distributions[16,18,40]. Then, the obtained fidelity distribution is fitted by the Gaussian distribution and extract its standard deviation. We find that just after preparing $|\downarrow\downarrow\rangle$, the state fidelity is only $98\%$ (Fig. 4c, g) and subsequent qubit controls do not decrease the state fidelity much ($< 2\%$) (Fig. 4d-f, 4h-j). This indicates that the imperfection of state preparation and measurement error removal contributes $\sim 1\text{-}2\%$ infidelity to the obtained state infidelities.

**Data availability:**

The data that support the findings of this study will be available from URL.

**Code availability:**

All codes used in this study are available upon reasonable request from the corresponding authors.

**Acknowledgements**

We thank the Microwave Research Group in Caltech for technical support. This work was supported financially by Core Research for Evolutional Science and Technology (CREST), Japan Science and Technology Agency (JST) (JPMJCR15N2 and JPMJCR1675), MEXT Quantum Leap Flagship Program (MEXT Q-LEAP) grant Nos. JPMXS0118069228, JST Moonshot R&D Grant Number JPMJMS2065, and JSPS KAKENHI grant Nos. 16H02204, 17K14078, 18H01819, 19K14640, and 20H00237. T.N. acknowledges support from JST PRESTO Grant Number JPMJPR2017.

**Author contributions**

A.N. and K.T. fabricated the device and performed the measurements. T.N. and T.K. contributed the data acquisition and discussed the results. A.S and G.S developed and supplied the $^{28}$silicon/silicon-germanium heterostructure. A.N. wrote the manuscript with inputs from all co-authors. S.T. supervised the project.

**Competing interests**

The authors declare that they have no competing interests.

**Additional information**

Correspondence and requests for materials should be addressed to A.N. or S.T.

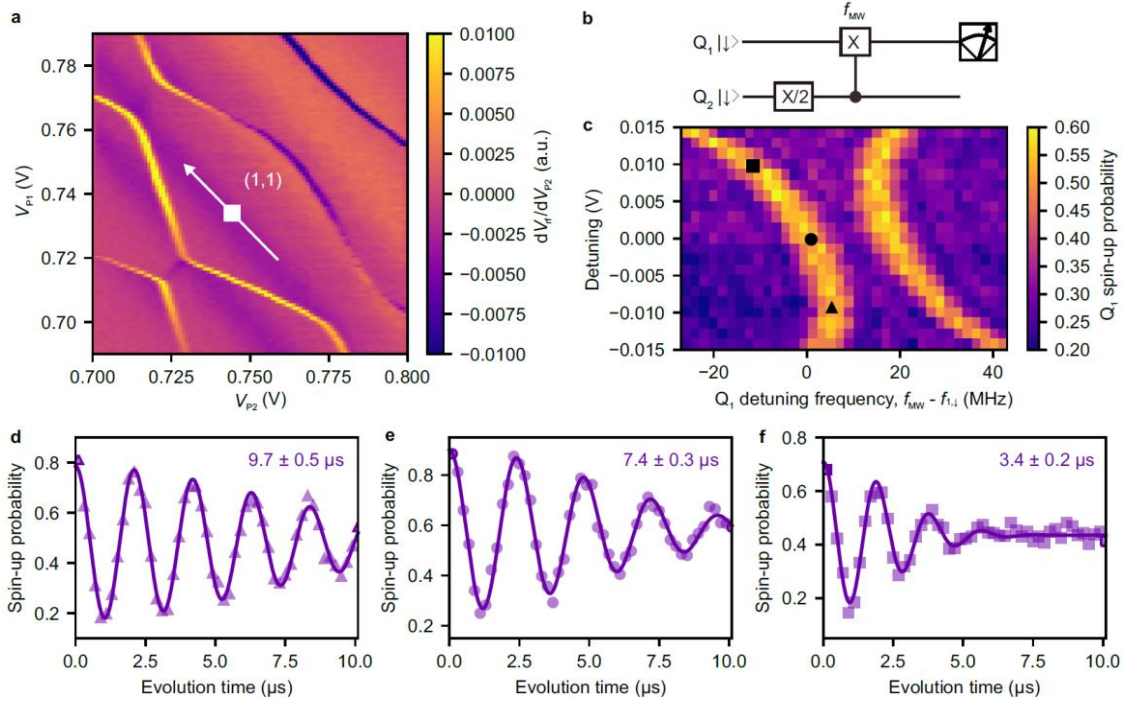

**Extended Data Figure 1. Detuning dependence of EDSR spectra. a,** Stability diagram around the (1,1) charge state. **b,** Quantum circuit for producing **c**. The microwave frequency of the $\pi$ CROT on $Q_1$ is changed to measure EDSR spectra. **c,** Detuning dependence of EDSR spectra of $Q_1$. The detuning axis and its origin are shown as the white arrow and square in **a**. Three black symbols show the conditions where the dephasing times $T_{2,1,\downarrow}^*$ shown in **d-f** are measured. **d-f,** Ramsey fringe of $Q_1$ when $Q_2$ is spin-down measured at the detuning $= -0.009$ V (**d**), $0$ V (**e**), and $0.009$ V (**f**). The integration time is 87 s for all of the traces. The errors in $T_{2,1,\downarrow}^*$ represent the estimated standard errors for the best-fit values. We observe longer (shorter) $T_{2,1,\downarrow}^*$ when the slope of the EDSR frequency against the detuning is smaller (larger), indicating the detuning charge noise limits $T_{2,1,\downarrow}^*$ at the charge-symmetry point where a finite slope exists due to the micromagnet-induced gradient field. A similar tendency is also observed in all the $T_{2,\mathrm{m},\sigma}^*$.

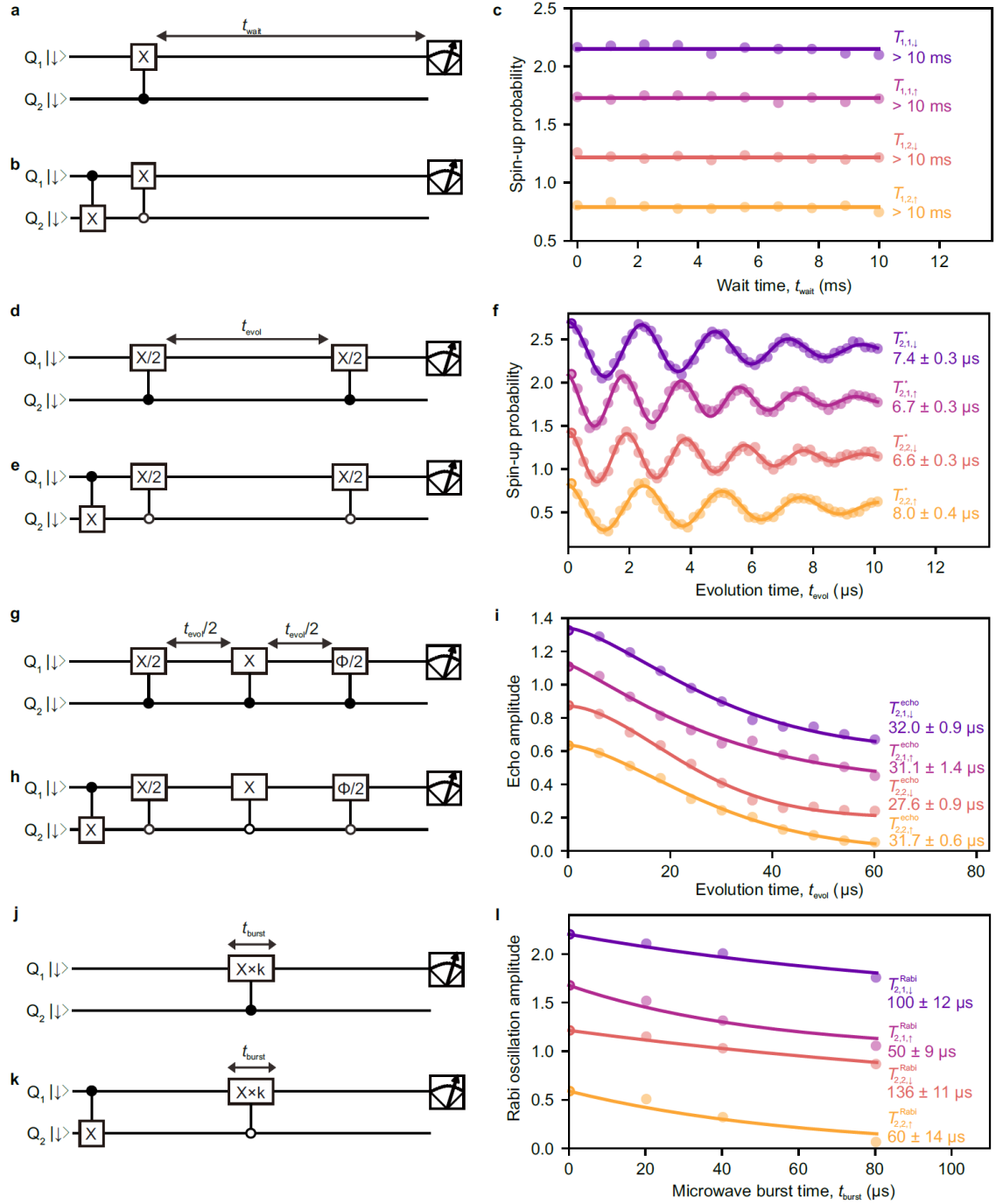

**Extended Data Figure 2. Qubits characterizations. a, b,** Sequences to measure spin relaxation times for $Q_1$ when $Q_2$ is spin-down, $T_{1,1,\downarrow}$ (**a**) and -up, $T_{1,1,\uparrow}$ (**b**). **c,** Spin-up probability as a function of the wait time. All of the traces do not show a decaying property indicating that spin relaxation is negligible for both qubits. The purple (magenta) curve is obtained using the sequence shown in **a** (**b**). The roles of $Q_1$ and $Q_2$ are swapped to measure the data for $Q_2$. Each trace is offset by 0.45 for clarity. All of the measurements are performed with $J = 18.85$ MHz and $f_R = 4.867$ MHz. **d, e,** Ramsey

sequences to measure dephasing times for $Q_1$, $T^*_{2,1,\downarrow}$ and $T^*_{2,1,\uparrow}$. **f,** Ramsey fringes of $Q_1$ and $Q_2$ fitted with Gaussian decaying oscillation functions. The integration time is 87 s for all of the traces. The errors represent the estimated standard errors for the best-fit values. Each trace is offset by 0.6 for clarity. **g, h,** Echo sequences to measure echo times for $Q_1$, $T^{\text{echo}}_{2,1,\downarrow}$ and $T^{\text{echo}}_{2,1,\uparrow}$. A phase of the final $\pi/2$ rotation is changed and the amplitude of the measured oscillation as a function of the phase is plotted in **i**. **i,** Echo amplitudes as a function of the evolution time. The exponent of the decay is 1.5, 1.2, 1.8, and 1.6 for $T^{\text{echo}}_{2,1,\downarrow}$, $T^{\text{echo}}_{2,1,\uparrow}$, $T^{\text{echo}}_{2,2,\downarrow}$, and $T^{\text{echo}}_{2,2,\uparrow}$. The errors represent the estimated standard errors for the best-fit values. Each trace is offset by 0.2 for clarity. **j, k,** Measurement of Rabi decay time for $Q_1$, $T^{\text{Rabi}}_{2,1,\downarrow}$, and $T^{\text{Rabi}}_{2,1,\uparrow}$. We measure Rabi oscillations by varying microwave burst time $t_{\text{MW}}$ from 0.01 μs to 0.41 μs with a separation of 0.01 μs. Rabi oscillations for longer $t_{\text{MW}}$ (offset by 20, 40, and 80 μs) are also measured and the amplitudes of the oscillations are plotted in **l**. **l,** Rabi oscillation amplitude as a function of the microwave burst time with decaying fits. The decay follows $R_{m,\sigma}(t) = \exp(-t/T^{\text{Rabi}}_{2,m,\sigma})W(t)$ where $W(t) = \left(1 + t^2 / \left(f_R(T^*_{2,m,\sigma})^2\right)^2\right)^{-1/4}$ represents the effect of dephasing[31]. From the fit, we extract the Rabi decay during a $\pi/2$ CROT as $D_{m,\sigma} = R_{m,\sigma}(t = 1/(4f_R))$. The errors represent the estimated standard errors for the best-fit values. Each trace is offset by 0.5 for clarity.

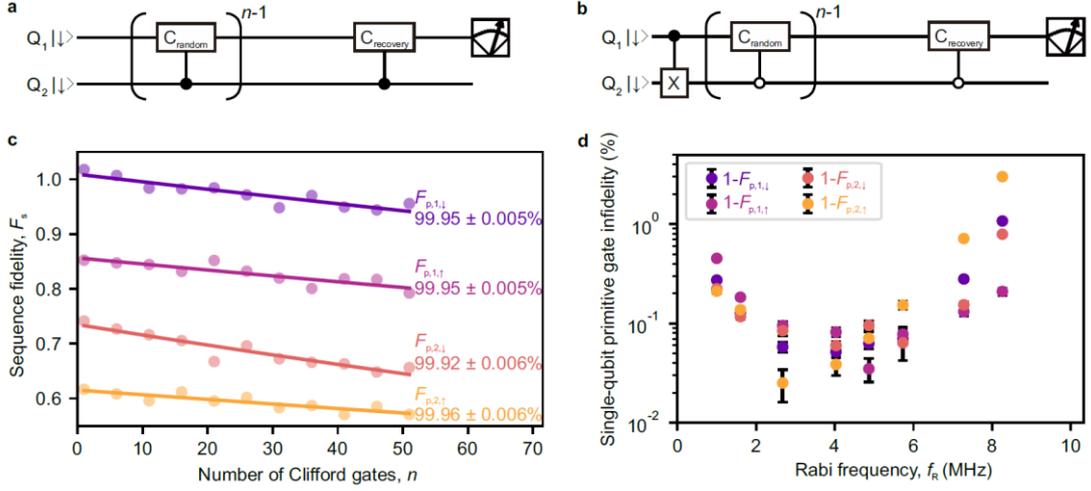

**Extended Data Figure 3. Single-tone single-qubit gate performance. a, b,** Quantum circuits of single-tone single-qubit Clifford-based randomized benchmarking for $Q_1$ when $Q_2$ is spin-down (**a**) and -up (**b**). **c,** Single-tone single-qubit primitive gate fidelities $F_{p,m,\sigma}$ assessed by the Clifford-based randomized benchmarking. The purple (magenta) curve is obtained using the sequence shown in **a** (**b**). The roles of $Q_1$ and $Q_2$ are swapped to measure the data for $Q_2$. $f_R = 4.867$ MHz and $J = 18.85$ MHz are used. Each trace is offset by $0.15$ for clarity. The uncertainty in the gate fidelities are obtained by a Monte Carlo method[4]. The obtained fidelities are consistent with those obtained in Fig. 2c as $F_{p,m} \sim F_{p,m,\downarrow} F_{p,m,\uparrow}$. **d,** Rabi frequency dependence of single-tone single-qubit primitive gate infidelities. Since the control qubit state is fixed in this measurement, the off-resonant rotation does not matter so that $f_R$ can be varied under a fixed $J$ of $32.0$ MHz. Therefore, the impact of $f_R$ on the single-qubit gate performance is assessed without involving the effect of $J$. We find that the fidelities depend on $f_R$ and the best values are obtained at $f_R = 2$-$5$ MHz. The uncertainty in the gate fidelities are obtained by a Monte Carlo method[4].

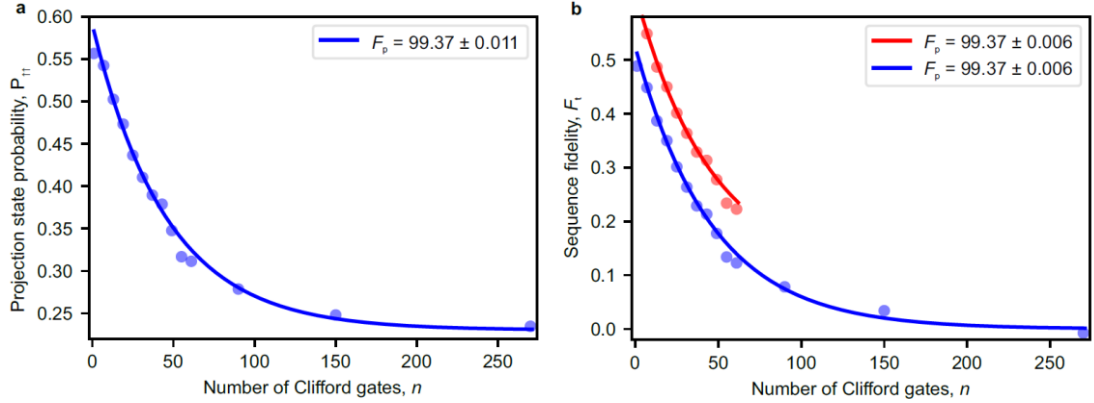

**Extended Data Figure 4. Two-qubit gate fidelity extraction. a,** Number of Clifford gates $n$ dependence of the projection state probability $P_{\uparrow\uparrow}$[4,16]. The ideal final state is spin-up for both qubits. To extract gate fidelity, we need to measure the saturation value of $P_{\uparrow\uparrow}$ with a large $n$ (Methods). The uncertainty in the gate fidelity is obtained by a Monte Carlo method[4]. **b,** Gate fidelity extraction from the sequence fidelity $F_t$. In addition to the data in **a**, we measure another data set where the final ideal state is spin-down for both qubits and then obtain $F_t$ as shown in blue (Methods). The saturation value of $F_t$ is almost zero ($F_t(271) = -0.007$) as expected. Gate fidelity extraction using only the data up to $n = 62$ is shown in red. The uncertainty in the gate fidelities are obtained by a Monte Carlo method[4]. The trace is offset by 0.1 for clarity. The obtained gate fidelities agree well with that obtained in the standard protocol in **a**. The uncertainty in the fidelity is larger in **a** due to the uncertainty of the saturation value of $P_{\uparrow\uparrow}$. $f_R = 5.732$ MHz and $J = 22.2$ MHz are used.

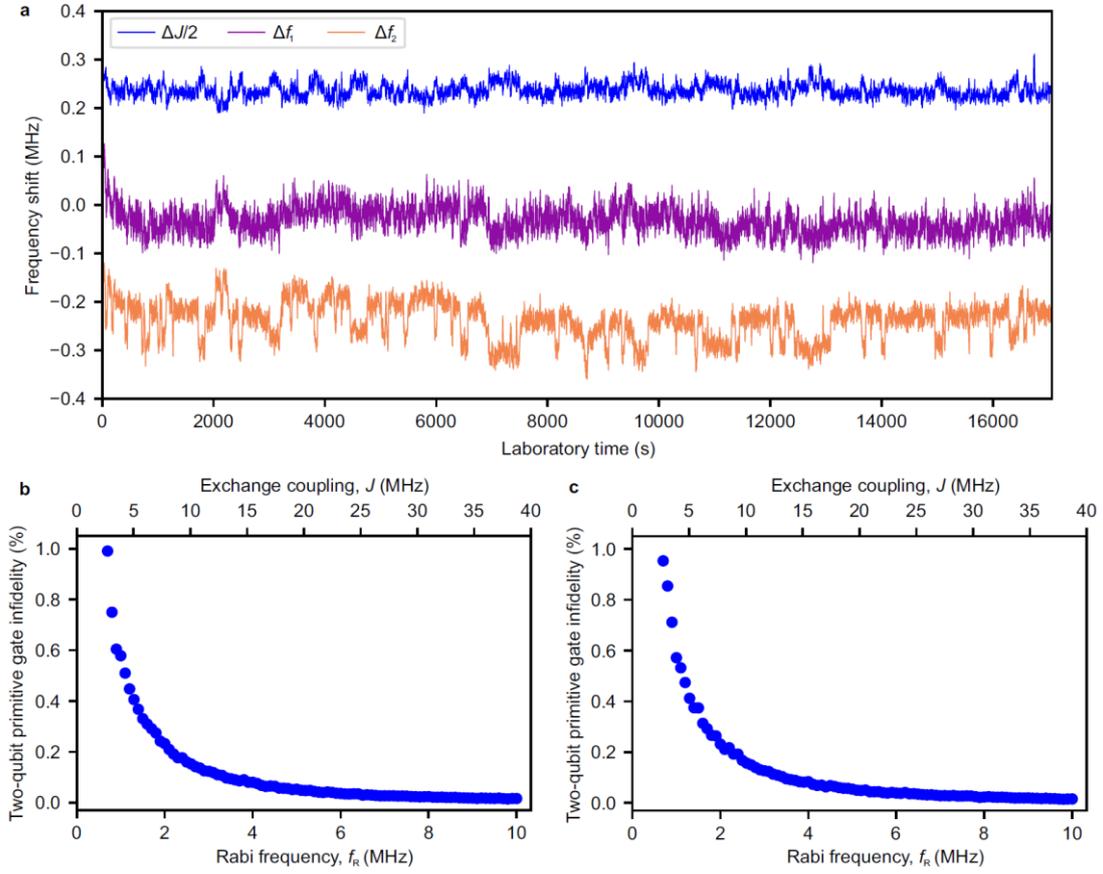

**Extended Data Figure 5. Estimation of two-qubit primitive gate infidelity by resonance frequency noise. a,** Time dependence of $\Delta J/2 = (\Delta f_{1,\uparrow} - \Delta f_{1,\downarrow})/2$ (blue), $\Delta f_1 = (\Delta f_{1,\uparrow} + \Delta f_{1,\downarrow})/2$ (purple), and $\Delta f_2 = (\Delta f_{2,\uparrow} + \Delta f_{2,\downarrow})/2$ (orange) extracted from repeated Ramsey fringe measurements (Methods). $J$ is fixed at 18.85 MHz. Each trace is offset by 0.25 MHz for clarity. Single-qubit frequency noises ($\Delta f_1$ and $\Delta f_2$) are larger than that of the exchange noise $\Delta J/2$. **b,** Simulation of a two-qubit primitive gate infidelity by the frequency noises obtained in **a** (Methods). **c,** Similar to **b** but the case with inserting an idle time for both qubits to remove the controlled-phase accumulation during the CROT when switching $J$ on and off[19,30].

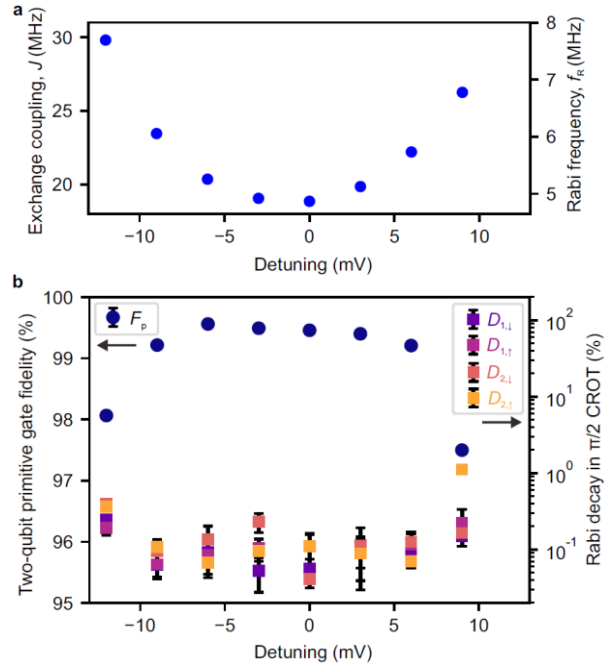

**Extended Data Figure 6. Detuning dependence of the two-qubit gate performance. a,** Detuning dependence of $J$. $J$ at the charge-symmetry point (detuning $=0$ mV) is $18.85$ MHz. **b,** Detuning dependence of the two-qubit primitive gate fidelity $F_\mathrm{p}$ (indigo circles) and the Rabi decay during the $\pi/2$ CROT (colored squares) obtained similarly to Fig. 1f. Around the charge-symmetry point, we reproducibly obtain $F_\mathrm{p}$ higher than $99\%$. In large positive and negative detuning, $F_\mathrm{p}$ sharply drops mainly due to the fast Rabi decay. The uncertainty in the gate fidelity is obtained by a Monte Carlo method[4]. The errors in the Rabi decay represent the estimated standard errors for the best-fit values.

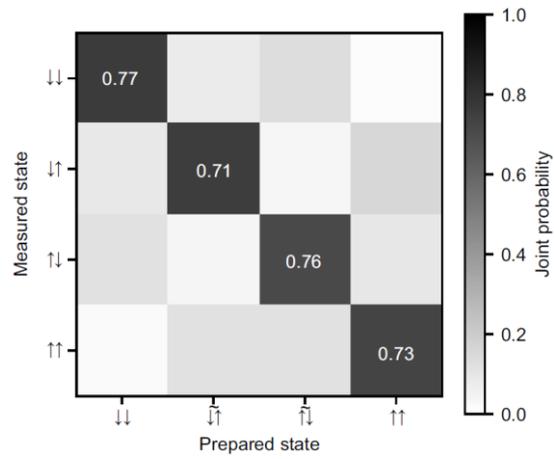

**Extended Data Figure 7. Measurement error calibration in state tomography.** Typical joint probabilities measured with preparing $|\uparrow\uparrow\rangle$, $|\widetilde{\uparrow\downarrow}\rangle$, $|\widetilde{\downarrow\uparrow}\rangle$, and $|\downarrow\downarrow\rangle$. At $J = 18.85$ MHz, $|\widetilde{\downarrow\uparrow}\rangle = 0.9995|\downarrow\uparrow\rangle + 0.0310|\uparrow\downarrow\rangle$.

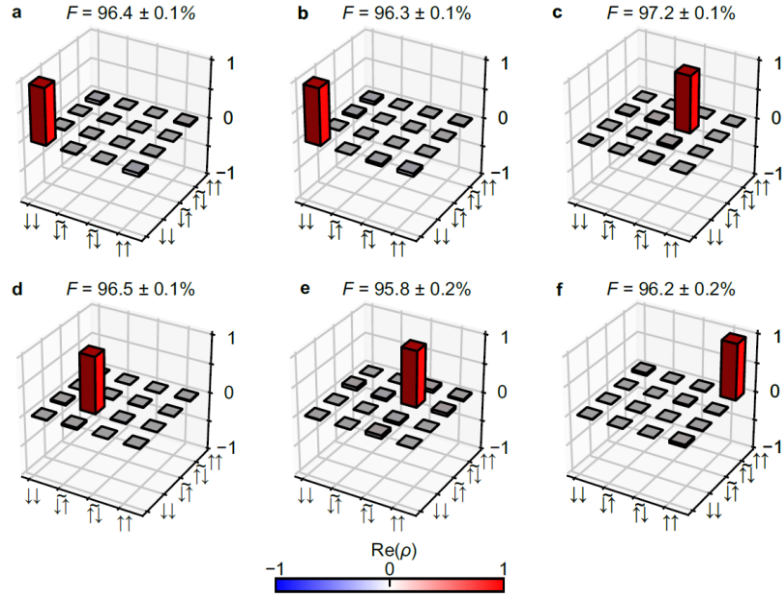

**Extended Data Figure 8. Output state of Deutsch–Jozsa algorithm and Grover search algorithm.**
**a-c,** Real part of the measured density matrix for the final output states for $f_0$ (**a**), $f_1$ (**b**), and $f_3$ (**c**) in the Deutsch–Jozsa algorithm (Fig. 4a). **d-f,** Real part of the measured density matrix for the final output states for $f_{10}$ (**d**), $f_{01}$ (**e**), and $f_{00}$ (**f**) in the Grover search algorithm (Fig. 4b). The absolute values of the matrix elements for the imaginary parts are less than $0.055$ (**a**), $0.056$ (**b**), $0.040$ (**c**), $0.111$ (**d**), $0.072$ (**e**), and $0.081$ (**f**). The uncertainty in the state fidelities $F$ are obtained by a Monte Carlo method[16,18,40].